\documentclass[sn-mathphys,Numbered]{sn-jnl}

\usepackage{graphicx}%
\usepackage{multirow}%
\usepackage{amsmath,amssymb,amsfonts}%
\usepackage{amsthm}%
\usepackage{mathrsfs}%
\usepackage[title]{appendix}%
\usepackage{xcolor}%
\usepackage{textcomp}%
\usepackage{manyfoot}%
\usepackage{booktabs}%
\usepackage{algorithm}%
\usepackage{algorithmicx}%
\usepackage{algpseudocode}%
\usepackage{listings}%
\usepackage{siunitx}%
\usepackage{subcaption}
\usepackage{tikz}
\usepackage{comment}
\usepackage{expl3}

\usepackage{soul,xcolor}
\setstcolor{red}

\theoremstyle{thmstyleone}%

\theoremstyle{thmstyletwo}%

\theoremstyle{thmstylethree}%

\begin{document}

\title[Article Title]{Kinetic Inductance and Jitter Dependence of the Intrinsic Photon Number Resolution in Superconducting Nanowire Single-Photon Detectors}

\author*[1,3,4]{\fnm{Roland} \sur{Jaha}}\email{r$\textunderscore$jaha01@uni-muenster.de}

\author[2,3,4]{\fnm{Connor A.} \sur{Graham-Scott}}

\author[2,3,4]{\fnm{Adrian S.} \sur{Abazi}}

\author[1]{\fnm{Wolfram} \sur{Pernice}}

\author[2,3,4]{\fnm{Carsten} \sur{Schuck}}

\author*[1]{\fnm{Simone} \sur{Ferrari}}\email{simone.ferrari@kip.uni-heidelberg.de}

\affil[1]{\orgdiv{Kirchhoff-Institute for Physics}, \orgname{University of Heidelberg}, \orgaddress{\street{Im Neuenheimer Feld 227}, \city{Heidelberg}, \postcode{69120 }, \country{Germany}}}

\affil[2]{\orgdiv{Department for Quantum Technology}, \orgname{University of Münster}, \orgaddress{\street{Heisenbergstraße 11}, \city{Münster}, \postcode{48149}, \country{Germany}}}

\affil[3]{\orgname{CeNTech - Center for Nanotechnology}, \orgaddress{\street{Heisenbergstraße 11}, \city{Münster}, \postcode{48149}, \country{Germany}}}

\affil[4]{ \orgname{SoN - Center for Soft Nanoscience}, \orgaddress{\street{Busso-Peus-Straße 10}, \city{Münster}, \postcode{48149}, \country{Germany}}}

\maketitle

\section*{Abstract}

The ability to resolve photon numbers is crucial in quantum information science and technology, driving the development of detectors with intrinsic photon-number resolving (PNR) capabilities. Although transition edge sensors represent the state-of-the-art in PNR performance, superconducting nanowire single-photon detectors (SNSPDs) offer superior efficiency, speed, noise reduction, and timing precision. Directly inferring photon numbers, however, has only recently become feasible due to advances in readout technology. Despite this, photon-number discrimination remains constrained by the nanowire's electrical properties and readout jitter.
In this work, we employ waveguide-integrated SNSPDs and time-resolved measurements to explore how the nanowire kinetic inductance and system jitter affect PNR capabilities. By analyzing the latency time of the photon detection, we can resolve changes in the rising edge of the detection pulse. We find that lower jitter as well as increased kinetic inductance enhances the pulse separation for different photon numbers and improves the PNR capability. Enhancing the kinetic inductance from 165\,nH to 872\,nH improves PNR quality by 12\%, 31\% and 23\% over the first three photon numbers, though at the cost of reducing the detector’s count rate from 165\,Mcps to 19\,Mcps. Our findings highlight the trade-off between PNR resolution and detector speed.

\section{Introduction}

Single photons have emerged as one of the most promising candidates for encoding and processing quantum information \cite{zhong2020quantum, madsen2022quantum, chen2021integrated}. This progress has been driven by significant advancements in the development of stable single-photon emitters \cite{aharonovich2016solid, zhai2022quantum} and high-performance single-photon detectors, broadening the scope of quantum technology applications. Photon number resolved detection plays a distinguished role in this regard, because it is a prerequisite for exploiting the full potential of linear optical quantum computation (LOQC, \cite{knill2001scheme}), boson sampling \cite{deng2023gaussian}, quantum enhanced imaging and sensing \cite{meda2017photon}, quantum Lidar \cite{cohen2019thresholded} and quantum enhanced communication \cite{becerra2015photon}. Therefore, the development of photon-number resolving detectors (PNRDs), which accurately reveal the photon number of a quantum state, has intensified over the recent years. So far, transition edge sensors (TESs) \cite{lita2008counting, calkins2013high} have been a preferred choice for PNRDs, achieving high distinguishability of large photon numbers \cite{eaton2023resolution}. However, their practical use in quantum technology applications is limited by the requirement for operation at millikelvin temperatures, long recovery times in the range of microseconds, and low timing resolution on the order of nanoseconds \cite{FUKUDA2024135}.

In contrast, superconducting nanowire single-photon detectors (SNSPDs) promise high efficiency ($>$\,98\,$\%$) \cite{reddy2020superconducting}, count rates in the GHz-regime \cite{craiciu2023high} and low timing jitter \cite{Korzh_2020} ($<$\,5\,ps). However, exploiting intrinsic PNR capabilities of SNSPDs has remained a challenge \cite{cahall2017multi} and previous work has instead focused primarily on quasi-PNR schemes that utilize temporal-multiplexing \cite{natarajan2013quantum} or spatial-multiplexing \cite{Gaggero:19, mattioli2016photon} to quantify the probability for a given light field to contain a certain number of photons. Accurately resolving photon numbers here requires a number of multiplexed detector elements significantly larger than the number of incoming photons, thereby reducing the likelihood of multiple photons reaching the same nanowire element of the SNSPD, which can be expressed as PNR quality \cite{jonsson2019evaluating}. Building on this approach, a hybrid spatiotemporal 100-nanowire element implementation achieved a PNR quality of more than 90\,$\%$ for photon numbers $n\leq 5$ \cite{cheng2023100}. 

Recently, there has been increasing interest in exploring the intrinsic properties of SNSPDs to resolve photon numbers to go beyond quasi-PNR schemes with a single nanowire detector  \cite{Nicolich_2019, sauer2023resolvingphotonnumbersusing}.
The core idea is to leverage the dependence of the SNSPD's turn-on dynamics on the number of incident photons, $n$. When $n$ photons strike the detector, they can create $n$ distinct resistive hotspots, contributing to an overall resistance $R(n)$. The slope of the rising-edge of the resulting output signal is then influenced by incident photon number, as it follows an approximate relationship $L_{\text{k}}/R(n)$, where $L_k$ is the detector's kinetic inductance \cite{cahall2017multi}. As more photons generate additional resistive spots, the overall resistance increases, which in turn modifies the slope of the signal’s rising edge, providing a way to infer the photon number. Leveraging this intrinsic effect, near-unitary PNR accuracy for photon numbers up to $n=4$ was demonstrated using a single detector connected to a large-inductance superconducting microstrip \cite{kong2024large}. 
Notably, large kinetic inductance here extends the rise time of the electrical output waveform, making variations in the rising edge originating from the absorption of different photon numbers easier to distinguish. 
However, while an increase in kinetic inductance improves photon number resolution, it also extends the output pulse duration thus reducing the achievable count rate and enhancing the timing jitter. This trade-off between the PNR capability of an SNSPD and its primary performance metrics requires careful consideration in implementations of quantum communication, information processing and sensing, which often require both reliable photon-number resolution as well as high timing accuracy and high count rates \cite{lita2022development}.

Here, we investigate how the intrinsic photon-number resolving capabilities depend on the electrical properties of the SNSPD. We evaluate the resulting trade-offs for waveguide-integrated SNSPDs (WI-SNSPDs \cite{pernice2012high, Ferrari2018}), which allow for more design freedom than conventional meander-shaped nanowires. Due to their traveling wave geometry optical modes are absorbed along the direction of propagation, resulting in high absorption efficiency per unit length for WI-SNSPDs \cite{5152999}, such that even short nanowires achieve appreciable detection efficiency. The waveguide-integrated detector configuration therefore not only offers the possibility for assessing a wide range of kinetic inductances but also holds great potential for achieving high detection rates in quantum photonic experiments. 

We here employ nine different WI-SNSPDs with kinetic inductances ranging from 165\,nH to 872\,nH for measuring latency time histograms between a laser trigger signal and the detector signal with a time-tagging device. The histograms can theoretically be modeled by a convolution of overlapping Gaussian functions, each centered at latency times that can be associated with specific photon numbers with a width representing the corresponding detector jitter. We find that the separation in latency time between the Gaussian functions follows a square-root dependence on kinetic inductance. Larger temporal separation, as well as reduced jitter, reduces the overlap of the Gaussian functions from minimal to maximal kinetic inductance, thus improving PNR quality by 12\,$\%$, 31\,$\%$ and 23\,$\%$ for single-photon, two-photon, and three-photon events, respectively. This improvement in PNR quality, however, significantly reduces the detector's maximally achievable count rate, dropping from 165\,Mcps to 19\,Mcps. This highlights the inherent trade-off between optimizing photon-number resolution and preserving high timing performance, where improvements in one aspect inevitably compromise the other. 
%

\section{Experimental Setup and Detector Fabrication} 

\begin{figure}[ht]
    \centering
\includegraphics[width=\linewidth]{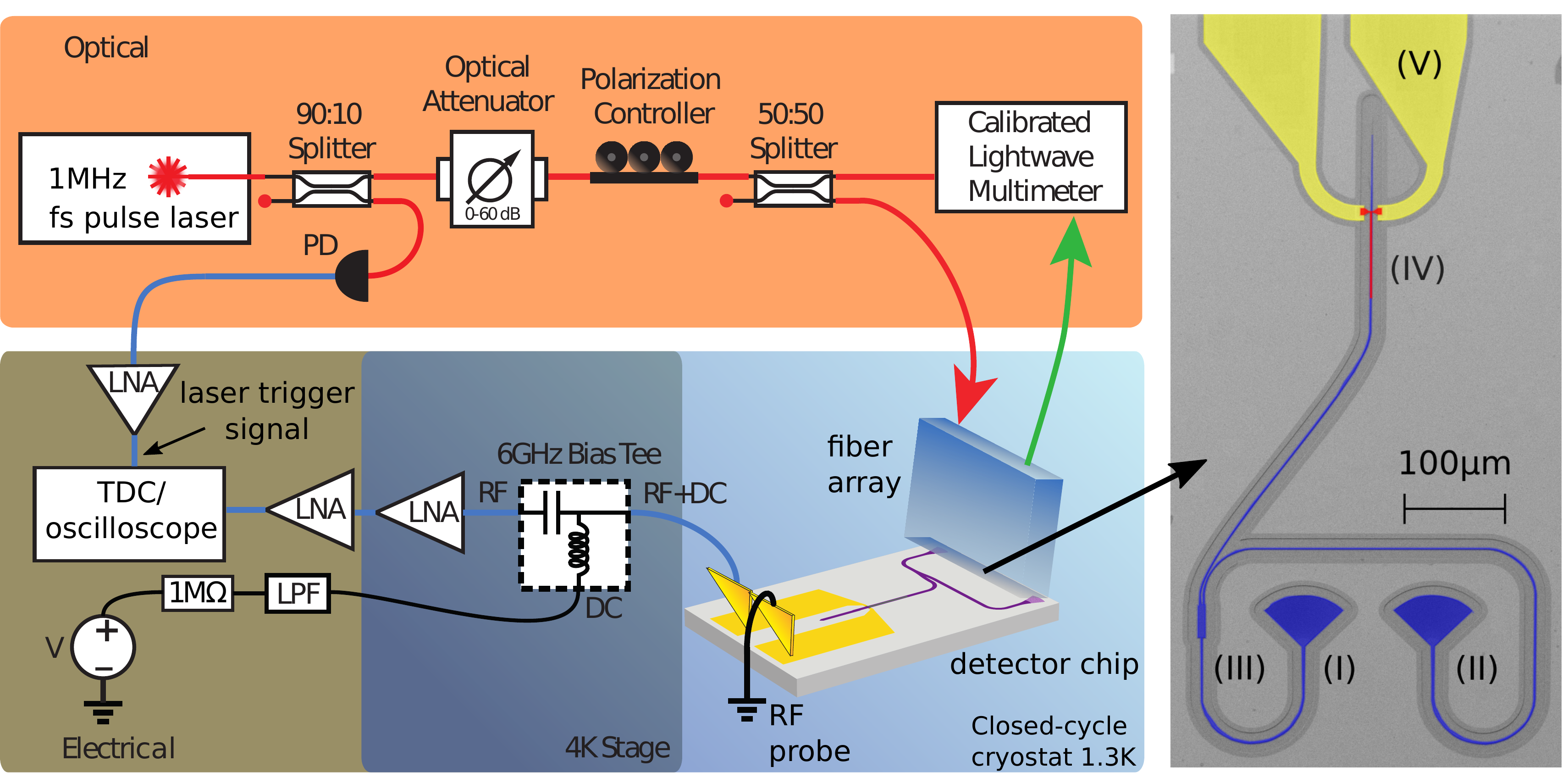}
    \caption[]{Left: Schematic measurement setup consisting of the optical input (orange), the electrical readout at low temperature (cyan) and at room temperature (brown) and the detector chip with optical and electrical access on a sample holder inside the cryostat (blue). Right: False color scanning electron micrograph of a single detector device integrated with grating couplers ((I) and (II)), nanophotonic waveguides (IV), MMIs (III) and electrode pads (V).}
    \label{fig:setup}
\end{figure}

We characterize nanophotonic chips with large numbers of WI-SNSPDs in a closed-cycle cryostat equipped with optical and electrical access, for delivering a calibrated photon stream to the detector and recording the response of the devices under test, respectively. The setup is depticted in Figure\,\ref{fig:setup}. Optical pulses of approx. 100\,fs duration are generated by a 1550\,nm wavelength laser (Elmo 780, Menlo Systems) with a fixed repetition rate of 1\,MHz (see orange box), to minimize artifacts such as afterpulses and AC biasing \cite{ferrari2019analysis}. The optical pulses are first split by a 90:10 fiber beam splitter, directing 10\,$\%$ of the light to a photodiode (PD) used for triggering the time-resolved measurement and the other 90\,$\%$ to a variable optical attenuator (VOA, HP8156A) that allows precise adjustment of the mean photon number reaching the SNSPD. After passing the attenuator, the optical pulses pass through a polarization controller and a 50:50 fiber beam splitter, which allows for measuring the photon flux reaching the cryostat using a calibrated lightwave multimeter (HP1635A). The optical signal is then coupled to the detector chip via a fiber array (red arrow).

The detector is current-biased using a voltage source meter (Keithley 2400) with a 1 M$\Omega$ series resistor. Two low-pass filters are used to attenuate any residual high-frequency noise from the source. The electrical response of the device under test is read out with a bias tee (Mini-Circuits ZFBT-GW6+), a cryogenic low-noise amplifier (Cosmic Microwave Technology CITLF3) cascaded to a room temperature low-noise amplifier (ZFL-1000LN+), connected to either an oscilloscope (Agilent MSO-X 91304A) or to a time-to-digital converter (TDC, Time Tagger X, Swabian Instruments), as depicted in the brown and blue boxes of Figure \,\ref{fig:setup}. The electrical signal, which is generated by the PD serves as a trigger signal for start-stop measurements in the time domain. It is amplified by another room-temperature LNA (ZFL-1000LN+) and provides a reference input for the TDC.

The detector chip with the waveguide-integrated SNSPDs is characterized inside a closed-cycle cryostat operating at 1.3\,K (blue box) and mounted on a low-temperature compatible 3D-nanopositioner (attocube ANP). Optical coupling from a fiber array to on-chip nanophotonic waveguides is realized by aligning the input grating couplers (I) on the chip, as shown in the scanning electron micrograph of Figure\,\ref{fig:setup}, to the single mode optical fibers in the array. The coupled light is split on-chip using a Multi-Mode Interferometer (MMI, (III)), where half of the light is directed to the output grating coupler (II) and the other half to the SNSPD (IV). The measurement of the optical power going to the input grating coupler and leaving the output grating coupler allows for the precise determination of the number of photons reaching the nanowire \cite{rath2016travelling}. The detector is connected to two gold electrodes (V), where electrical access for supplying bias current and readout is realized using a radio-frequency (RF) probe.

The waveguide-integrated detectors in this study are fabricated using a top-down approach from a NbTiN/$\text{Si}_3\text{N}_4$/Si$\text{0}_2$ on Si layer stack. The superconducting NbTiN thin film, approximately 6\,nm thin, is deposited in a room temperature process \cite{wolff2020superconducting}. After patterning alignment markers and electrode pads (Au) in a lift-off process, nanowires are patterned in electron-beam lithography using CSAR resist and CF4-based reactive ion etching. Subsequently, waveguides, MMIs and grating couplers are fabricated in another electron-beam lithography step, employing CSAR resist as a mask for fluorine-based 
etching. Each detector consists of a single hairpin-shaped nanowire and is electrically connected to the on-chip electrode pads.

\newpage

\section{Characterization of SNSPD timing performance}

\begin{figure}[ht]
\centering
\includegraphics[width=\linewidth]{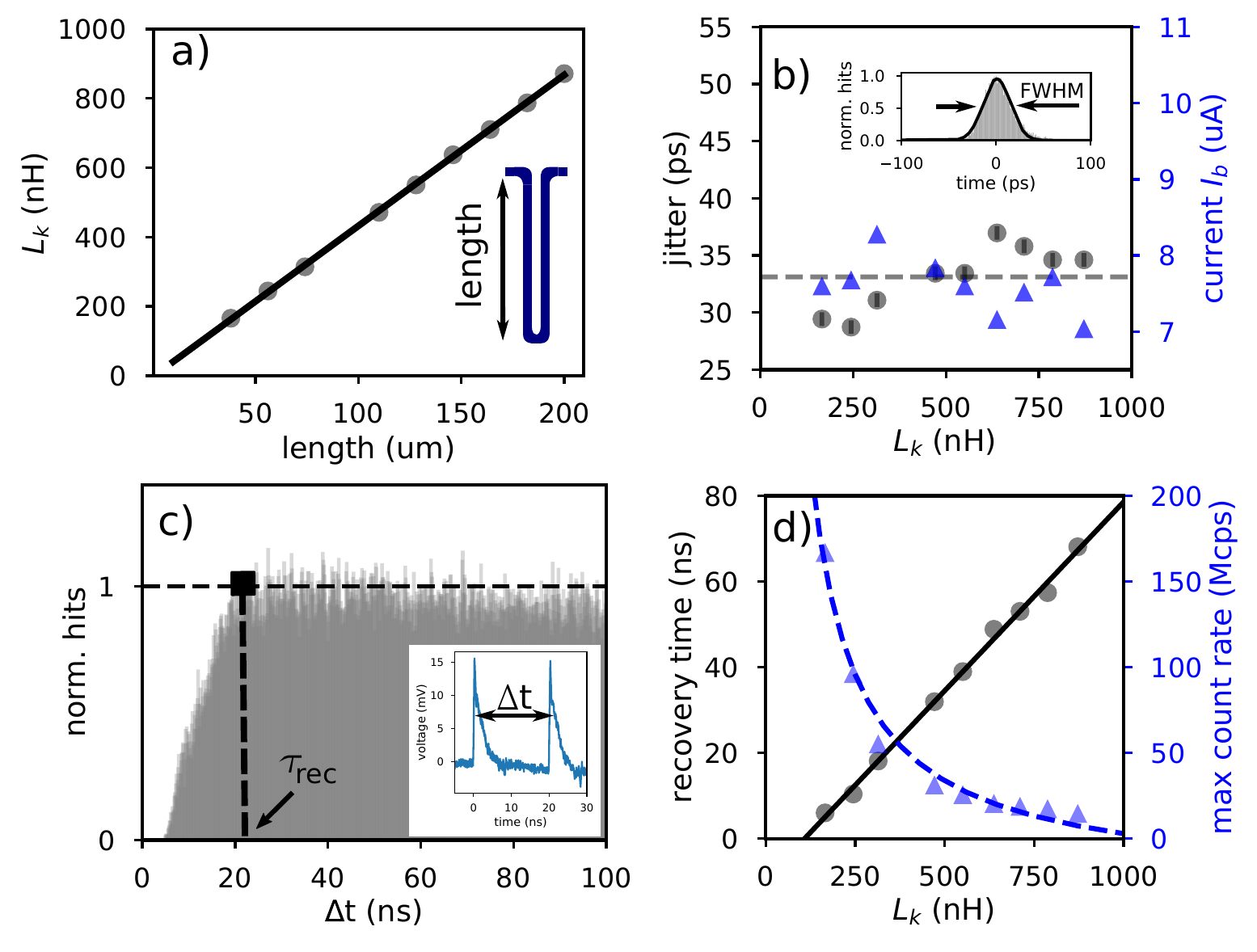}
\caption{Characterization of 100\,nm wide detectors with varying lengths. a) Relationship of the detector length and kinetic inductance. b) Dependence of the jitter on the kinetic inductance. c) Example of an inter-arrival time histogram to calculate the recovery time of a detector. d) Recovery time and maximum count rate as a function of the kinetic inductance.}
\label{fig:snspd_characterization}
\end{figure}

We realize detectors with different kinetic inductances by varying the length $l$ of the nanowire, while keeping the width constant at 100\,nm. Figure \ref{fig:snspd_characterization}a) shows the dependence of the kinetic inductance on the length of the wire, where we extract $L_k = Z_0{\tau}_{\text{dec}}$ from the decay time ${\tau}_{\text{dec}}$ of recorded electrical signal traces and the load impedance of the readout circuitry $Z_0$\,=\,50\,$\Omega$. We observe a linear increase of the kinetic inductance, following the relationship $L_k\propto l$, in the range from 165\,nH to 872\,nH. 

For the purpose of this work on PNR, we focus on two key performance metrics: detector timing resolution (jitter) and maximum count rate (recovery time). To evaluate the timing resolution, we record latency time histograms between the detector signal and the pulsed laser trigger signal and calculate the full width at half maximum (FWHM) of the Gaussian fit to these histograms. The results are illustrated in Figure \ref{fig:snspd_characterization}b), which shows the extracted jitter values as a function of the kinetic inductance. All histograms as well as subsequently reported PNR measurements are recorded at detector bias currents of $I_b=0.8\cdot I_c$, where $I_c$ represents the critical current of the detector. At these bias currents, the detectors operate in a high signal-to-noise ratio regime, where both dark count events and latching occur with low probability.
As shown in Figure \ref{fig:snspd_characterization}b, all devices exhibit a similar critical current of $I_b = 7.6\pm 0.3$\,$\mu$A, with variations primarily originating from fabrication differences across the detectors. These variations affect the timing jitter, as the jitter is inversely proportional to the bias current \cite{10.1063/1.4817581}. On average, the detectors exhibit a jitter of $\bar{j}=(33\pm 2)\,ps$.

To determine the maximally achievable count rate, we illuminate the detector with a continuous-wave (CW) laser and use an oscilloscope to collect an inter-arrival time (IAT) histogram \cite{10.1063/1.3691944}. This histogram captures the time delays $\Delta$t between subsequent detection events, as shown in Figure \ref{fig:snspd_characterization}c. From the IAT histograms, we extract the recovery time ${\tau}_{\text{rec}}$, which is the time required for the detector to regain its full detection efficiency after an event. The maximally achievable count rate $R^{\text{max}}_C$ is then calculated as the inverse of the recovery time: $R^{\text{max}}_C = 1 / {\tau}_{\text{rec}}$. 

The results for the recovery times and corresponding maximum count rates for all detectors are depicted in Figure \ref{fig:snspd_characterization}d). We observe that the recovery time increases linearly with kinetic inductance, ranging from ${\tau}_{\text{rec}} = 5.99$\,ns (at 165\,nH) to ${\tau}_{\text{rec}} = 68.11$\,ns (at 872\,nH). Consequently, the maximally achievable count rates exhibit an inverse proportionality with kinetic inductance, decreasing from 165\,Mcps (at 165\,nH) to 19\,Mcps (at 872\,nH). 

Our characterization revealed a linear correlation between kinetic inductance and device length across the investigated range of detector geometries. Additionally, we observe an inverse relationship between kinetic inductance and count rate. Within the examined range and experimental parameters, kinetic inductance only shows marginal effects on timing jitter (see Supplementary Material \ref{sup:electrical_jitter}).

\section{Results and Discussion}

\subsection{PNR of a single detector}

\begin{figure}[ht]
\centering
\includegraphics[width=\linewidth]{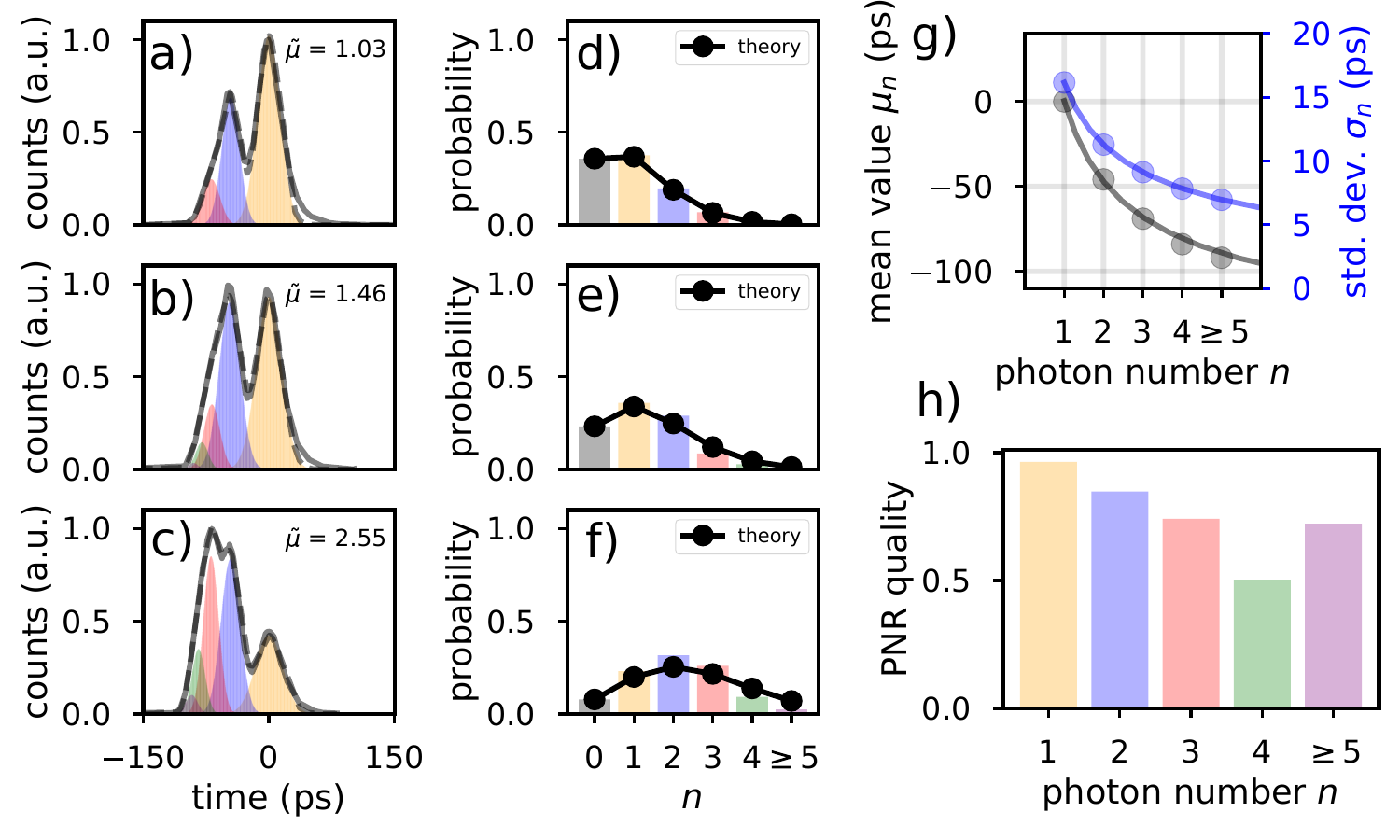}
\caption[]{Intrinsic photon number resolution of an SNSPD with width\,=\,100\,nm and length\,=200\,$\mu$m and 872\,nH kinetic inductance. a)-c) Latency time histograms at three different effective mean photon numbers ${\tilde{\mu}}$ with fitted Gaussian distributions representing discrete photon numbers $n$. d)-f) Respective photon count statistics reconstructed from the histogram distributions. g) Mean values $\mu_n$ and standard deviation $\sigma_n$ of each fitted Gaussian function for the histogram measured at ${\tilde{\mu}}$\,=\,2.66. h) Corresponding bar plot displaying the PNR quality of each photon number.}
\label{fig:histograms_att}
\end{figure}

Information about the number of photons absorbed by a nanowire can be extracted from the rising edge of the output waveform produced by the SNSPD. Here we investigate the photon-number resolving capabilities of our nanowire detectors by measuring the latency time between the laser trigger signal and the detection signal for fixed voltage trigger levels of the time tagger. Figures \ref{fig:histograms_att}(a)-(c) show histograms for three exemplary effective mean photon numbers incident on a detector with nanowire length l\,=\,200\,$\mu$m and width w\,=\,100\,nm, corresponding to a kinetic inductance of 872\,nH. We extract the effective mean photon numbers using the equation $\tilde{\mu} = -\text{ln}(1-CR/RR)$, that takes into account the detection efficiency ${\eta}_{\text{eff}} = 1-e^{-\tilde{\mu}} = CR/RR$, where $CR$ is the rate of detected events and $RR$ denotes the repetition rate of the laser. 

At low effective mean photon numbers, $\tilde{\mu} = 1.05$, the histogram shows distinct peaks that can be well-fit by overlapping Gaussian distributions. We associate each Gaussian with specific photon detection events: single-photon ($n=1$, yellow), two-photon ($n=2$, blue), three-photon ($n=3$, red), and smaller peaks corresponding to higher photon numbers. At low effective mean photon numbers, the contribution from single-photon events is dominant. As the effective mean photon number, $\tilde{\mu}$, increases, the contributions associated with the detection of larger photon numbers become more pronounced: the two-photon peak ($n=2$) is most noticeable at $\tilde{\mu} = 1.46$, and the three-photon peak ($n=3$) emerges clearly at $\tilde{\mu} = 2.55$. Importantly, we observe that the histograms show a decreasing separation between the Gaussian distributions underlying each peak as the photon number increases, and beyond $n=5$, the peaks become too closely spaced to resolve them unambiguously.

In Figures \ref{fig:histograms_att}(d)-(f), we compare the photon statistics extracted from the measured histograms with the discrete Poisson distribution $Q(n)={\tilde{\mu}}^n e^{-\tilde{\mu}}/n!$, which accurately describes the behavior of the coherent light field supplied in this experiment.
Each bar plot represents the area under the Gaussian fit corresponding to a specific photon number, colored to match Figures \ref{fig:histograms_att}(a)-(c). The $n=0$ contribution is determined by quantifying the number of laser pulses with no recorded photon detection event, following the equation: ${CR}_{n=0} = RR-CR$. The black dots in Figures \ref{fig:histograms_att}(d)-(f) represent the expected values for discrete Poisson distributions of effective mean photon numbers $\tilde{\mu} = 1.05, 1.46, 2.55$, respectively. The good agreement between the expected and the measured photon number distributions, following Poissonian statistics, validates that our detectors effectively operate as true photon-number-resolving (PNR) devices.

In Figure \ref{fig:histograms_att}(g), we plot the mean values ${\mu}_n$ and standard deviations ${\sigma}_n$ of the fitted Gaussian distributions from the histogram measured at an effective mean photon number $\tilde{\mu}=2.66$. As the photon number $n$ increases, the latency time difference $\Delta$t between two neighboring peaks decreases, causing the Gaussian distributions to overlap more strongly. Specifically, the timing difference $\Delta t_{1,2}=\mu_1-\mu_2$ of the single-photon and two-photon Gaussian distributions is 49\,ps, whereas it decreases to $\Delta t_{2,3}=\mu_2-\mu_3=23$\,ps for the time difference between the two-photon and three-photon distributions, and further to $\Delta t_{3,4}=\mu_3-\mu_4=15$\,ps. 

This trend can be modeled using the relationship $\mu_n \propto \sqrt{n^{-1}}$, which also effectively captures the relationship between the rise time of the detection signal and the photon number \cite{Nicolich_2019}. A similar pattern is observed for the standard deviations ${\sigma}_n$ of the Gaussian peaks, which decreases from 16.3\,ps ($n=1$) to 7.0\,ps ($n\geq 5$). The narrowing of the Gaussian function for larger $n$ results from the scaling of the rise time with the photon number, $\tau_{\text{rise}} \propto \sqrt{n^{-1}}$. Consequently, because the eletrical jitter is proportional to the rise time \cite{WAHL1963247, zhao2011intrinsic}, the standard deviation exhibits the same scaling behavior with the photon number $n$: $\sigma_n \propto \sqrt{n^{-1}}$.

Based on the model proposed by \cite{morais2024precisely}, we present a similar method to assess the photon-number resolution capability of a detector in time-domain measurements. The PNR quality is determined by the probability $P^m_n$, which represents the likelihood of correctly assigning photon number m to a detection event. To calculate this value, we first define the threshold boundaries in the latency time histogram as the intersection points $[t_m,t_{m+1}]$ between adjacent Gaussian distributions corresponding to different photon numbers n. For each Gaussian distribution $G_n(t)$, the integral over the normalized distribution within the interval $[t_m,t_{m+1}]$ is computed, representing the contribution of photon number n within this range.
The probability $P^m_n$ is then obtained by normalizing this integral through division by the total integral of all Gaussian distributions within the same interval:

\begin{equation}
    P^m_n = \frac{\int_{t_m}^{t_{m+1}}G_n(t)dt}{\sum_{n} \int_{t_m}^{t_{m+1}}G_n(t)dt}
\label{equation_1}
\end{equation}

In Figure \ref{fig:histograms_att}(h), we present the values for $P^m_n$ for the histogram captured at an effective mean photon number $\tilde{\mu}=2.66$. We observe that the PNR quality decreases with the photon number from 0.96 ($n=1$) to 0.49 ($n=4$).

\subsection{Influence of kinetic inductance and timing jitter on the PNR capability}

\begin{figure}[h]
\centering
\includegraphics[width=\linewidth]{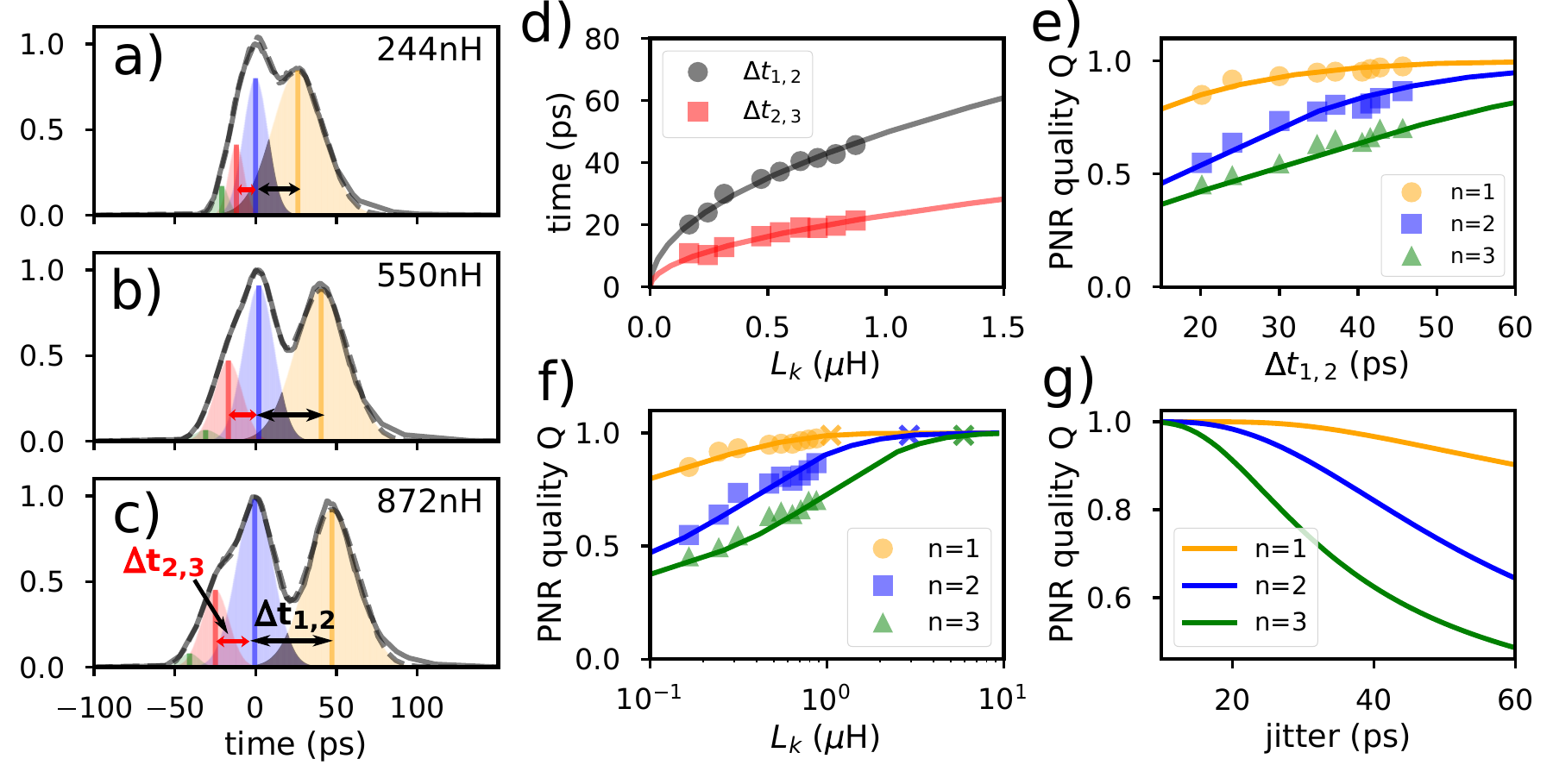}
\caption[]{a)-c) Latency time histograms measured at $\tilde{\mu}= 1.5$ of three different detectors with $L_k$\,=\,244\,nH, 550\,nH, and 872\,nH. d) $\Delta t_{1,2}$ and $\Delta t_{2,3}$ as functions of the kinetic inductance showing a square-root relationship. e) Relationship between the PNR quality and the time difference, derived from a theoretical model including four overlapping Gaussian functions. f) Corresponding dependence of the PNR quality on the kinetic inductance. g) PNR quality as a function of the jitter for a detector with a kinetic inductance of 880\,nH.}
\label{fig:histogram_different_devices}
\end{figure}

Expanding on the characterization of our detectors and the methodology introduced earlier, we now explore how the PNR capability of the detector is influenced by its kinetic inductance and timing jitter. In Figures \ref{fig:histogram_different_devices}(a)-(c), we present the latency time histograms for three detectors with different kinetic inductances, all measured at an effective mean photon number of $\tilde{\mu}= 1.5$. For better comparability, the peak position for $n=2$ is shifted to the center across all histograms. 

We perform a multi-Gaussian peak fit of the histogram, where the first peak corresponds to $n=1$ (yellow), the second to $n=2$ (blue), the third to $n=3$ (red), and the fourth to $n=4$ (green), with the peak positions indicated by vertical lines. This fit accurately captures the histogram data, as the probability of recording a count for $n\geq 5$ is less than 2\,$\%$ under the given photon illumination. 

We observe a noticeable enhancement in the separation of peaks with the increase of the device kinetic inductance. For $L_k$\,=\,872\,nH, $\Delta t_{1,2}$ is 47\,ps, resulting in the smallest overlap of the peaks. In comparison, the device with an intermediate kinetic inductance exhibits a $\Delta t_{1,2}$ of 41\,ps, whereas the one with the lowest kinetic inductance shows a $\Delta t_{1,2}$ of only 26\,ps.

We repeat this procedure for different detectors and plot $\Delta t_{1,2}$ and $\Delta t_{2,3}$ as a function of the kinetic inductance. The results are shown in Fig.\ref{fig:histogram_different_devices}(d). The data is consistent with a square-root relationship between the time difference and the kinetic inductance, with $\Delta t_{1,2}$ ranging from 21\,ps to 47\,ps, and $\Delta t_{2,3}$ ranging from 11\,ps to 21\,ps. This square-root dependence is further observed in the relationship between the rise time and the kinetic inductance (refer to Section \ref{sup:electrical_jitter} of the Supplementary Material), suggesting that the increased rise time leads to an enhanced separation between photon number peaks.

Hence, higher timing separation of the peaks can be achieved by increasing the kinetic inductance. However, with the range of kinetic inductance used in this experiment, full peak separation cannot be observed, making it challenging to completely eliminate overlap and to achieve unitary PNR quality. To address this, we theoretically evaluate the dependence of the PNR quality factor $Q$ on the time difference, aiming to estimate the kinetic inductance required for full peak separation.

Our approach employs a theoretical model that represents the detector response in the time-bin domain as the sum of four Gaussian functions, 
\begin{equation}
    \sum^{4}_{n=1} A_n \exp \left( -\frac{(t - \mu_n)^2}{2 \sigma_n^2} \right),
\end{equation}
where the Gaussians $G_n(t)$ correspond to specific photon numbers. Each Gaussian is defined by its amplitude $A_n$, mean value $\mu_n$, and standard deviation $\sigma_n$, capturing the distinct characteristics of each photon number state. In order to estimate $\sigma_n$ and $\mu_n$, we use the $1/\sqrt{n}$ relationship with the photon number (see Figure \ref{fig:histograms_att}(g)). The amplitudes $A_n$ are determined by the Poisson distribution $Q(n)$ and the resulting probability of detecting $n$ photons at a given effective mean photon number $\tilde{\mu}$. Additional details on the model can be found in Section \ref{supp:four_gaussian} of the Supplementary Material.

For the model we need to initialize the effective mean photon number $\tilde{\mu}$ and the standard deviation $\sigma_1$ of the Gaussian function corresponding to the first photon number. In order to adequately fit our measured data, these values are chosen based on the experimental conditions. We set the effective mean photon number to $\tilde{\mu}$\,=\,1.5 and calculate $\sigma_1$ using the measured average jitter across all devices $j_1 = 33$\,ps (see Figure \ref{fig:snspd_characterization}(b)), by $\sigma_1 = j_1 / 2\sqrt{2\text{ln}2}$. We perform a numerical sweep of ${\Delta t}_{1,2}$ and for each value of ${\Delta t}_{1,2}$, we calculate the PNR quality for the first three photon numbers according to Eq.\,\ref{equation_1}. Figure \ref{fig:histogram_different_devices}(e) shows the simulated model results for the PNR quality as solid lines, plotted alongside the measured values for $n=1$ (yellow), $n=2$ (blue) and $n=3$ (green). We observe a sigmoidal dependence of the PNR quality with ${\Delta t}_{1,2}$, where unitary PNR quality can be achieved for large  ${\Delta t}_{1,2}$. The model aligns well with the measured data, with only minor deviations, likely due to variations in jitter and an effective mean photon number across devices. Overall, the close agreement between the measured and simulated results indicates that our model effectively captures the PNR quality behavior across the range of fabricated kinetic inductances.

To determine the minimal kinetic inductance required to achieve full separation of the Gaussian peaks, we use the experimentally determined relationship between ${\Delta t}_{1,2}$ and kinetic inductance:  $\Delta t_{1,2}\propto \sqrt{L_k}$ (see Figure \ref{fig:histogram_different_devices}(d)). We derive the curve of Figure \ref{fig:histogram_different_devices}(f) which shows the dependency of the PNR quality with the kinetic inductance for $n=1$, $n=2$ and $n=3$. The curves reveal that increasing the kinetic inductance consistently drives the PNR quality closer to unity. We can use the model to extrapolate to larger kinetic inductances and estimate the minimum kinetic inductance required to achieve a PNR quality $Q$ exceeding 0.99, corresponding to  complete peak separation for the first three photon numbers: 1.2\,$\mu$H for $n=1$, 3.5\,$\mu$H for $n=2$, and 7.0\,$\mu$H for $n=3$. 

An important trade-off revelaed by our experimental results is that by increasing the kinetic inductance from 165\,nH to 872\,nH, the PNR quality across the first three photon numbers increases, with improvements of 12\,$\%$ for $n=1$, 31\,$\%$ for $n=2$, and 23\,$\%$ for $n=3$. However, this enhancement comes at the cost of a reduced maximum count rate at which the detector can operate. Specifically, the count rate drops nearly tenfold, from 165\,Mcps for the detector with the lowest kinetic inductance to 19\,Mcps for the one with the highest kinetic inductance (see Figure \ref{fig:snspd_characterization}(d)).

This intrinsic trade-off between PNR quality and detector count rate can be mitigated if the system timing jitter is reduced. Lower jitter leads to less overlap between peaks, thereby improving PNR quality. Using our model, we are able to quantitatively describe the influence of the system jitter on the PNR quality. For this analysis, we keep ${\Delta}t_{1,2}$ constant at 47\,ps, which corresponds to a kinetic inductance of 880\,nH, and sweep the jitter of the first peak from 10\,ps to 60\,ps. Figure \ref{fig:histogram_different_devices}(g) shows model data for the first three photon numbers, highlighting a sigmoidal relationship that underscores the impact of jitter on PNR quality. Notably, with jitter at 10\,ps, the PNR quality exceeds 0.99 for the first three photon numbers. This result confirms that reducing system jitter is essential for optimizing PNR quality without compromising detector performance. Several strategies exist for lowering system jitter. To minimize electrical jitter, adopting cryogenic amplifiers is highly effective, as they significantly improve the signal-to-noise ratio of the detection signal \cite{cahall2018scalable}. Additionally, jitter reduction can be achieved by increasing the critical current of the nanowire, which can be effectively optimized by adjusting the stoichiometry of the superconducting NbTiN film \cite{Zichi:19}.

\section{Conclusion}

In this work, we demonstrated the significant impact of kinetic inductance and system jitter on the intrinsic PNR capabilities of waveguide-integrated SNSPDs. We introduced a theoretical model to quantitatively describe how the intrinsic PNR capability is linked to these two key factors. Our model indicates that reducing system timing jitter decreases the overlap between peaks corresponding to different photon numbers, thus enhancing the PNR quality. Additionally, the model reveals a sigmoidal dependence of PNR quality on the timing separation between photon number peaks. This timing separation can be significantly increased by raising the kinetic inductance, as our experimental results demonstrate a square-root dependence between timing separation and kinetic inductance. For the range of kinetic inductances from 165\,nH to 872\,nH, the PNR quality improved by 12\,$\%$, 31\,$\%$ and 23\,$\%$ over the first three photon numbers, respectively. However, our work also highlights an intrinsic trade-off in the pursuit of resolving higher photon numbers. As the kinetic inductance increases, the recovery time of the detectors also increases, leading to a significant reduction in the detector count rate, from 165\,Mcps to 19\,Mcps. This trade-off underscores the limitations of relying solely on increased kinetic inductance to enhance photon-number resolution, particularly for applications in quantum sensing or quantum communication, where both high PNR capability and a large dynamic range are critical. However, with the continuous improvements in timing jitter in recent years through cryogenic amplifiers and optimization of the film quality, this compromise becomes less restrictive. As a result, SNSPDs hold great potential to emerge as strong competitors to transition-edge sensors. SNSPDs not only offer the ability to resolve higher photon numbers but also outperform TESs in all the other key performance metrics.

\section*{Acknowledgements}
This work has been supported by German Federal Ministry of Education and Research (BMBF) through the PhoQuant Project. We acknowledge the support from the European Union's Horizon 2020 Research and Innovation Action under Grant Agreement No. 899824 (FET-OPEN and SURQUID). C.S. acknowledges support from the Ministry for Culture and Science of North Rhine-Westphalia (421- 8.03.03.02-130428).

\bibliography{bibliography}
\bibliographystyle{apsrev4-1}

\newpage

{\noindent\Large\textbf{Supplementary Material}}

\newcounter{supsection}
\renewcommand\thesupsection{S\arabic{supsection}}
\renewcommand{\thefigure}{S\arabic{figure}}
\setcounter{figure}{0}

\phantomsection
\refstepcounter{supsection} 
\section*{\thesupsection\ Kinetic-inductance dependence of the rise time and the electronic jitter of waveguide-integrated SNSPDs}
\label{sup:electrical_jitter}
Electronic jitter and the rise time of a detector's voltage response signal are closely linked. Wahl \cite{WAHL1963247} proposed a model that quantifies the contribution of electronics to timing jitter in solid-state detectors, which has been shown to apply to SNSPDs as well \cite{zhao2011intrinsic}. According to this model, the electronic noise jitter can be expressed as:
\begin{equation}
    j_{elec}=\frac{\delta V}{SR}
\end{equation}
where $\delta V$ represents the full width at half maximum (FWHM) of the amplitude noise in the readout circuit, and SR is the slew rate of the voltage response pulse’s rising edge. The slew rate is proportional to the pulse amplitude and inversely proportional to the rise time, establishing an inverse relationship between electronic jitter and rise time. 
As the rise time increases with the square-root of the length $l$ \cite{Nicolich_2019}, this further reinforces the dependence of timing jitter on the kinetic inductance.
\begin{equation}
    j_{elec}\propto\frac{\delta V}{V_{peak}}\tau_{rise}\propto\frac{\delta V}{V_{peak}}\sqrt{L_{k}}
\end{equation}
where $V_{peak}$ is the maximum voltage response signal peak amplitude \cite{phdthesis}.

In time-bin photon number resolving measurements, an increase in timing jitter leads to greater overlap between the Gaussians corresponding to different photon sensitivities. However, as the overlap decreases with an increase in kinetic inductance, the corresponding rise in jitter presents a trade-off for this application. It is therefore crucial to ensure that the increase in jitter remains minimal, so it does not significantly broaden the Gaussians, which could otherwise impair the PNR resolution. We therefore investigate the effect of the kinetic inductance on jitter and rise time for the $L_k$ range under investigation.
\begin{figure}[H]
\centering
\includegraphics[width=.7\linewidth]{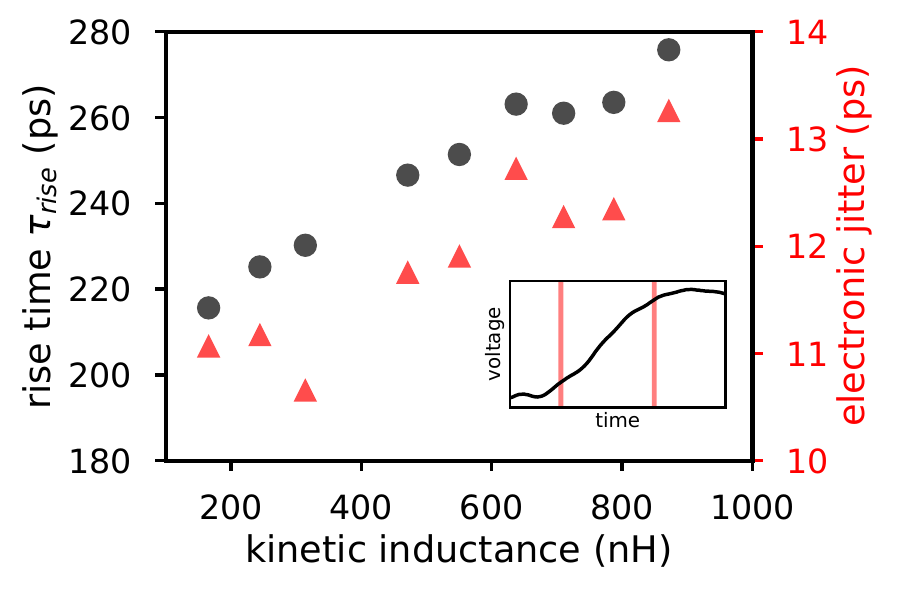}
\caption[]{Dependency of the rise time and electronic jitter with the kinetic inductance}
\label{fig:rise_time_kinetic_inductances}
\end{figure}
The results are shown in Fig. \ref{fig:rise_time_kinetic_inductances}, where we confirm the dependencies mentioned above. Additionally, we observe that the increase in jitter within the investigated range is minimal, and its detrimental effect on PNR quality at high kinetic inductances can be neglected. 
%

\phantomsection
\refstepcounter{supsection} 
\section*{\thesupsection\ Multi-peak model for fitting the time-bin detector response for PNR resolution}
\label{supp:four_gaussian}

In this section, we present a detailed explanation of the model used to describe the dependence of the PNR quality factor $Q$ on kinetic inductance. The current model is designed to resolve photon numbers in the range of $n=1$ to $n=4$, but it can be extended to accommodate higher photon number resolutions. According to our model, the measured histogram is fitted as the sum of four Gaussian peaks: 
\begin{align*}
    \sum^{4}_{n=1} G_n(t) = \sum^{4}_{n=1} A_n \exp \left( -\frac{(t - {\mu}_n)^2}{2 {\sigma}_n^2} \right).
\end{align*}
where each Gaussian peak, corresponding to a specific photon number $n$ is characterized by a standard deviation ${\sigma}_n$, and amplitude $A_n$ and a mean value in the time domain ${\mu}_n$.

\begin{figure}[H]
\centering
\includegraphics[width=\linewidth]{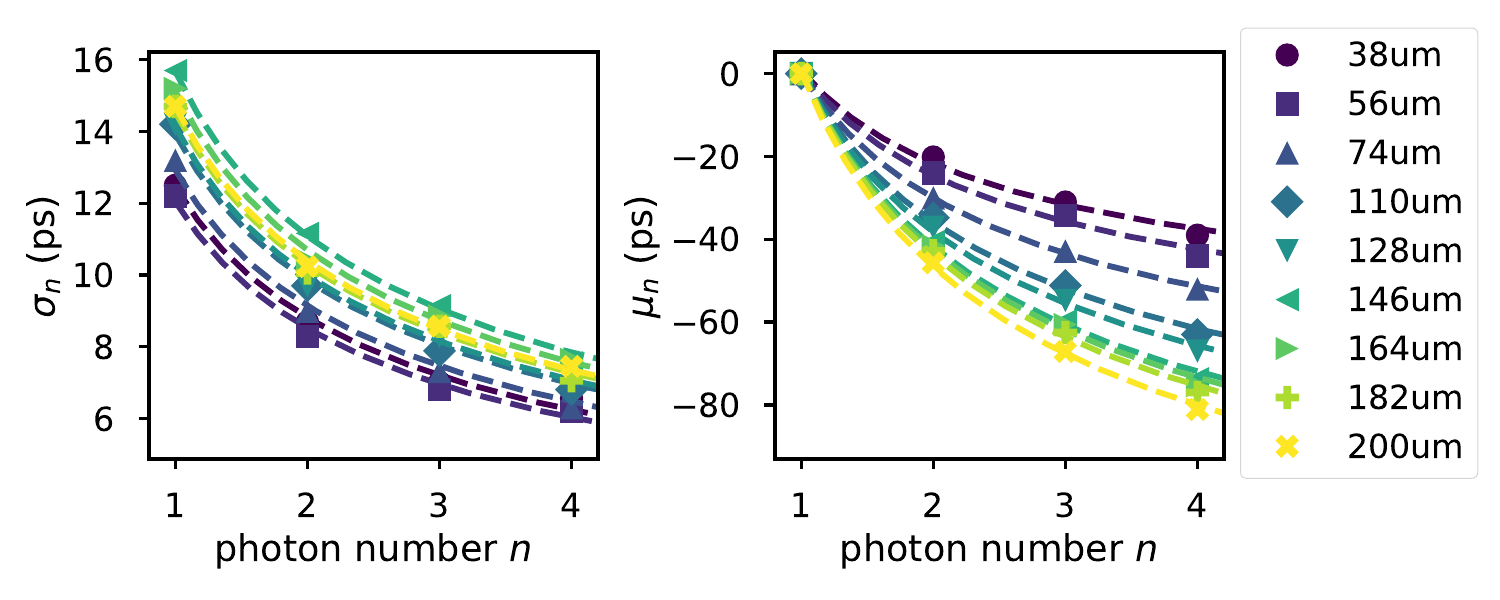}
\caption[]{Dependency of the width ${\sigma}_n$ and peak positions ${\mu}_n$ on photon number across devices with varying kinetic inductances. The $1/\sqrt{n}$ behavior is observed for both parameters across all measured devices.}
\label{fig:jitter_mu_vs_photon_number}
\end{figure}

We extract the standard deviation ${\sigma}_1$ for $n=1$ from our experimental data (Fig. \ref{fig:snspd_characterization}b). Using this value, we calculate the standard deviations ${\sigma}_n$ for higher photon numbers $n$ using the empirical relation:
\begin{align*}
{\sigma}_{n} = {\sigma}_{1} / \sqrt{n},
\end{align*}
as determined in (Fig. \ref{fig:jitter_mu_vs_photon_number}a).
The validity of our empirical relation can be confirmed by the dependence of the rise time $\tau_{rise}$ on the photon number, which follows:
\begin{align}
    \tau_{rise}\propto\frac{1}{\sqrt{n}}
\end{align}\cite{Nicolich_2019}. Because the electrical jitter is linearly proportional to the rise time (see \ref{sup:electrical_jitter}), it exhibits the same scaling behavior with the photon number.
The amplitudes $A_n$ represent the amplitudes of the Gaussian curves for each photon number $n$. They are calculated using the relative contributions $R_n$, which quantify the relative contributions of each photon number to all detection events, following Poisson statistics.

Each relative contribution $R_n$ corresponds to the area under the Gaussian curve $G_n(t)$ for photon number $n$. The area of a Gaussian is related to its amplitude $A_n$ and standard deviation ${\sigma}_n$ through the equation:
\begin{align*}
R_n = \int_{-\infty}^{\infty}G_n(t) = \sqrt{2\pi} A_n {\sigma}_n 
\end{align*}
Given $R_n$ (the relative contribution) and ${\sigma}_n$ (the standard deviation), $A_n$ is obtained as: 
\begin{align*}
A_n = \frac{R_n}{{\sigma}_n \sqrt{2\pi}}
\end{align*}

The mean values $\mu_n$ in the time domain, corresponding to the photon number $n$ in your model, are calculated using the model for the SNSPDs turn-on dynamics proposed by Nikolich et al. \cite{Nicolich_2019}, which follows an inverse square-root dependence on the photon number. Because we fix the mean value of the first peak to $\mu_1 = 0$, we calculate the remaining mean values using $\Delta t_{1,2}$:

\begin{align*}
    \mu_n = 
\begin{cases}
    0 & n = 1 \\
    \mu_{n-1} + \frac{\Delta t_{1,2}}{\sqrt{n-1}} & n\geq 2 \\
\end{cases}
\end{align*}

With all necessary variables defined, the model can fit the detector's response in the time-bin domain and be used to explore the dependence of PNR resolution on jitter and kinetic inductance.

\phantomsection
\refstepcounter{supsection} 
\section*{\thesupsection\ PNR on single-strip detectors}
\label{bowtie}

In our study, we investigate the dependence of photon number resolution (PNR) quality on the kinetic inductance of superconducting nanowire single-photon detectors (SNSPDs) by utilizing detectors of varying lengths. 

In this section, we aim to demonstrate that variations in detector length do not significantly affect the distribution of photon absorption along the nanowire, which typically occurs within the first few micrometers \cite{FerrariSchuckPernice+2018+1725+1758}. As a result, the localization of photon absorption has a negligible impact on the PNR response. Instead, we focus on the role of kinetic inductance, showing that the observed effects are primarily electrical in nature rather than related to absorption.

\begin{figure}[H]
\centering
\includegraphics[width=.8\linewidth]{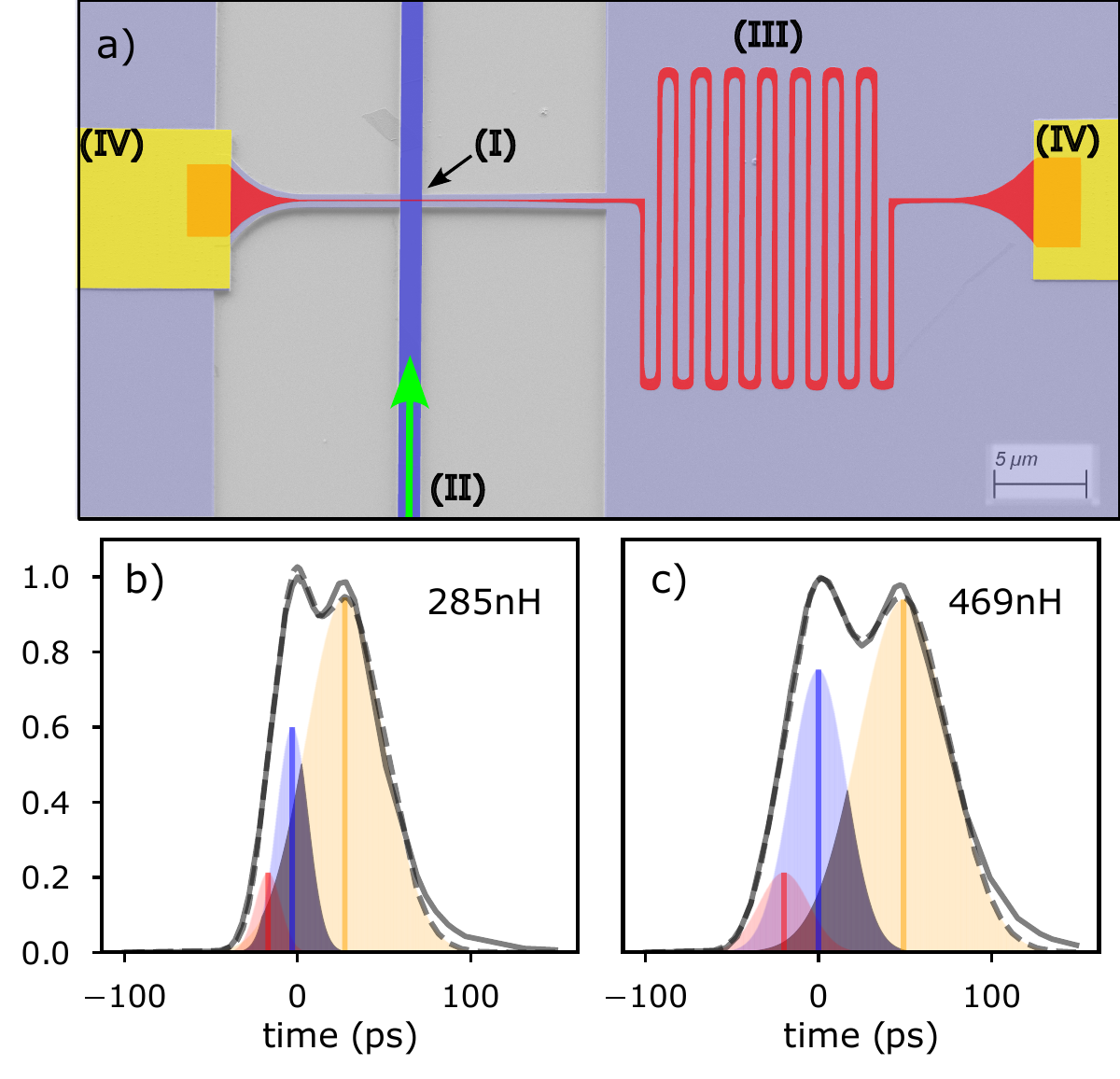}
\caption[]{(a) False-colored SEM image of the single-strip detector. The NbN detector consists of a straight section (I) oriented perpendicular to the waveguide (II), featuring a very small active area. A meandered section (III), made from the same superconducting film, is added in series with the detector to create different kinetic inductances. Electrical contact is established via gold contact pads (IV). (b,c) Latency time histogram for two detectors with different additional inductance.}
\label{fig:PNR_bowtie}
\end{figure}

To isolate the influence of kinetic inductance, we utilize bowtie-shaped nanowire detectors, which consist of a short, narrow strip crossing the waveguide. Due to their small cross-section, these devices are inherently inefficient, and photonic crystal cavities are often employed to enhance efficiency \cite{Muenzberg2018, Vetter2016}. However, in this work, we deliberately retain the low efficiency of the detectors and test the PNR quality by integrating an external inductor onto the chip. The configuration is depicted in Figure \ref{fig:PNR_bowtie} (a), where the active area responsible for photon absorption remains unchanged. The only variation between the devices lies in their additional kinetic inductance (indicated with III in Figure \ref{fig:PNR_bowtie} (a)).

Our measurements reveal that the PNR quality is strongly influenced by the kinetic inductance. The devices exhibit overall kinetic inductances of 285\,nH and 469\,nH, respectively and are biased at about 80\% of the critical current. As shown in Fig. \ref{fig:PNR_bowtie} (b,c).

We find that higher kinetic inductance results in improved peak separation in the PNR histograms. Specifically, the peak separation decreases from 48 ps for the higher kinetic inductance device to 27 ps for the lower inductance device, leading to a reduction in PNR quality. This result confirms that the PNR quality is dictated by the kinetic inductance of the device and not by the spatial distribution of resistive hotspots generated during photon absorption.

These findings are further supported by measurements on other detectors with different additional inductance.

\end{document}